\documentclass[preprint,showpacs,preprintnumbers,amsmath,amssymb]{revtex4}

\usepackage{graphicx}
\usepackage{dcolumn}
\usepackage{bm}

\begin{document}


\title{Geophysical test of the universality of free-fall}

\author{Sachie Shiomi}
 \affiliation{Mechanical
Measurement Laboratory, Measurement Standards \& Technology
Division, Center for Measurement Standards, Industrial Technology
Research Institute, Hsinchu, Taiwan 30011 R.O.C.}


\begin{abstract}
We point out that the universality of free-fall can be tested by
observing surface-gravity changes of the Earth. The Earth's inner
core is weakly coupled to the rest part of the Earth by mainly
gravitational forces. If there were a violation of the universality
of free-fall, because of their different chemical compositions
and/or of different mass fractions of binding energies, the inner
core and the rest part of the Earth would fall at different rates
towards the Sun and other sources of gravitational fields. The
differential acceleration could be observed as surface-gravity
effects. By assuming a simple Earth model, we discuss the expected
surface-gravity effects of violations of the universality and
experiments to search for such effects by using superconducting
gravimeters. It is shown that the universality can be tested to a
level of 10$^{-9}$ using currently operating superconducting
gravimeters. Some improvements can be expected from combinations of
global measurements and applications of advanced data analyses.
\end{abstract}

\pacs{04.80.Cc, 91.10.Pp} \keywords{the equivalence principle, the
universality of free-fall, superconducting gravimeters}

\maketitle

\section{\label{st:introduction}Introduction}

The universality of free-fall, stating that every material (point
mass) in a gravitational field falls at the same rate, is accepted
as one of the most fundamental principles in modern physics; the
Newtonian mechanics and Einstein's general relativity are based on
this principle. All experiments performed so far support the
universality. However, theories towards the unification of the four
forces (strong, weak, electromagnetic and gravitational
interactions) typically introduce new interactions and predict
violations of the principle
\cite{Lee1955,Fujii1971,Fayet1986,Taylor1988,Damour1994,Dimopoulos1996}.
These theories predict new Yukawa-potential type interactions
between putative charges, such as baryon number, isospin and
electrostatic energy density, which are functions of proton and
neutron numbers in elements. This prediction implies chemical
composition-dependence of free-fall. Also, many alternative metric
theories of gravitation predict violations of the universality of
free-fall of massive bodies \cite{Nordtvedt1968}. Therefore, testing
the universality at a high sensitivity is expected to make a
breakthrough in the current understanding of physics. Because there
are many unknown factors in those theories, no precise prediction
has been made for the magnitudes of the putative violations; the
universality should be tested as precisely as possible. Also,
because the characters of the putative interactions are uncertain,
the universality has to be tested for various putative charges,
using different kinds of test bodies, at different ranges. In
addition, to confirm experimental results, it should be tested by at
least two different experimental methods. Considering the importance
of testing the universality and the necessity of variety in
experimental approaches, we present a new method of testing the
universality. So far, it has been tested by various experiments
(e.g. \cite{Fischbach1999}) and it is verified to a level of
$10^{-13}$ by laboratory E{\"{o}}tv\"{o}s-type experiments
\cite{Smith2000} and by Lunar Laser Ranging (LLR)
\cite{Williams2004}.

The merits of the proposed method may be as follows: (1) this is the
first application of the superconducting gravimeters, whose
performance has been verified to be sensitive and stable in
geophysical studies, to tests of the universality, (2) this is the
first attempt to use Earth's interior as test bodies, and (3) unlike
most of the experiments for testing the universality, this method
has a potential to test the universality of free-fall of
gravitational self-energy.

We present the concept of this method in section \ref{st:Concept},
the theory in section \ref{st:Equation of motion} and the expected
sensitivity in section \ref{st:sensitivity}.

\section{\label{st:Concept}The concept}

In this method, the test bodies are the solid inner core and the
rest part of the Earth (the liquid outer core, the mantle and the
crust). They are in free fall in the gravitational field mainly due
to the Sun. If there were a violation of the universality of
free-fall, there would be differential acceleration between the
inner core and the rest part of the Earth towards the Sun (see
section \ref{st:Equation of motion} for detail). The differential
acceleration can be searched for by measuring surface-gravity
changes. Relative gravity changes can be measured with high
resolution by using superconducting gravimeters. They have been used
to search for translational motions of Earth's inner core (the
Slichter triplet \cite{Slichter1961})(e.g.
\cite{Smylie1992,Courtier2000,Rosat2003,Rosat2004,Sun2004}). It is
shown that they are capable of measuring long period surface-gravity
effects. We attempt to apply the geophysical tool to test the theory
of gravitation.

In the test bodies, there are mainly two kinds of differences that
could result in the differential acceleration due to violations of
the universality. One is the difference in their chemical
compositions (the chemical composition-dependent effect), which
implies the difference in proton and neutron numbers in the elements
that compose the test bodies. The solid inner core mainly consists
of iron and nickel, and its density is approximately $\rho_{ic}$
$\simeq$ 13000 kg m$^{-3}$ \cite{Anderson1989}, while the rest part
of the Earth is mainly made of lighter elements such as silicon
oxides, except the liquid outer core (which mainly consists of heavy
elements and whose density is slightly less than the inner core
($\rho_{oc}$ $\simeq$ 11000 kg m$^{-3}$ \cite{Anderson1989})). The
average density of the rest part, including the outer core, is
approximately 5400 kg m$^{-3}$.

The other is the difference in the fractions of gravitational
self-energy in the test bodies (about $-$3.7 $\times$ 10$^{-11}$ and
$-$ 4.2 $\times$ 10$^{-10}$ for the inner core and the rest part,
respectively). Testing the universality of free-fall of
gravitational self-energy can be viewed as a test of the strong
equivalence principle.

The chemical composition-dependent effect can be tested using
laboratory-size test masses. The effect of gravitational self-energy
appears significant only in massive test bodies.

\section{\label{st:Equation of
motion}Equation of motion of the inner core}

\begin{figure}
\includegraphics[width= 5 cm]{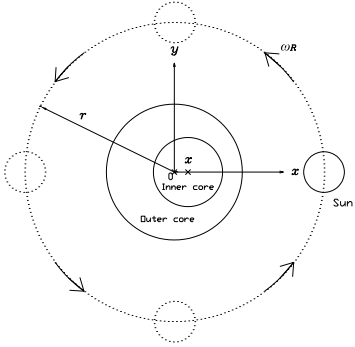}\caption{A schematic
cross section of the assumed configuration (not drawn to scale). The
Earth is fixed and the Sun goes around the Earth in the circular
orbit (radius $r$) at the angular frequency $\omega_R \approx 2
\pi$(24\ h)$^{-1}$. The inclination of Earth's rotation axis to
ecliptic is assumed to be 0$^\circ$ for simplicity. The $x$-$y$
plane is set on the Earth's equatorial plane. The Earth's inner core
oscillates along the x-axis. The gravitational influence from
Earth's mantle and crust (not drawn) is negligible because of
Newton's shell theorem (see text).} \label{Fig:Earth-model}
\end{figure}

We assume a simple configuration as shown in figure
\ref{Fig:Earth-model}. The Earth is revolving around the Sun (with
angular frequency $\omega_R$) in a circular orbit with radius $r$.
The Earth's rotation axis is perpendicular to the ecliptic plane,
namely the inclination of the rotation axis by about 23.3$^\circ$ is
not considered. The $x$-$y$ plane is set on the Earth's equatorial
plane with its origin at the center of figure of the rest part of
the Earth.

The Earth's interior can be classified into four parts: the solid
inner core, the liquid outer core, the mantle and the crust. We
assume that the solid inner core is a homogeneous sphere (with
density $\rho_{ic}$ and radius $r_{ic}$ $\simeq$ 1.22 $\times$
10$^{6}$ m \cite{Anderson1989}) and it is enclosed in the spherical
liquid outer core (with the average density $\rho_{oc}$ and outer
radius $r_{oc}$). We do not consider any deformations (such as tidal
or rotational deformations). We assume that the mantle and the crust
are spherical shells with uniform densities, and their centers of
figures are coincident; their gravitational influence on the inner
core is negligible due to Newton's shell theorem. The Coriolis
acceleration splits oscillations of the inner core into a triplet of
periods (the Slichter triplet \cite{Slichter1961}). However, for
simplicity, we assume that the inner core oscillates only along the
$x$-axis (with angular frequency $\omega_0$) and we ignore the
Coriolis acceleration. We do not consider any electromagnetic
effects.

The effective viscosity of the outer core $\eta$ is not well
determined and estimates from various methods vary from $\sim$
10$^{-3}$ Pa s to $\sim$ 10$^{12}$ Pa s \cite{Secco1995}. The
Reynolds number ($\equiv \rho_{oc} v r_{ic}/ \eta$, where $v$ is the
velocity of the inner core) is less than unity when $\eta$ is larger
than $\sim$ 10$^7$ Pa s and the amplitude of oscillations $X$
($\equiv$ $v/\omega_0$) is nominally 1 m. We assume that the
friction between the inner core and the outer core is proportional
to the velocity of the inner core.

Under these assumptions, the $x$-component of the equation of motion
of the inner core (mass $m_{ic}$ = 4 $\pi r_{ic}^3 \rho_{ic}$/3
$\approx$ 1.0 $\times$ 10$^{23}$ kg) can be written approximately in
a form of so-called damped forced oscillation:
\begin{equation}
\ddot{x} \approx - \omega_0^2x - 2k\dot{x} + \Delta a \cos \omega_R
t\label{eq:eq_of_motion}
\end{equation}
where
\begin{eqnarray}
\omega_0^2 &\approx& \frac{4}{3} \pi G \frac{\rho_{ic} -
\rho_{oc}}{\rho_{ic}} \rho_{oc} \approx 4.7 \times 10^{-7}
\rm{s^{-2}} \approx \{2 \pi (2.5 \hspace{5 pt}\rm{h})^{-1}\}^2
\label{eq:omega_0^2}\\
2k &\equiv& \frac{6 \pi r_{ic} \eta}{m_{ic}} \approx 2.3 \times
10^{-16} \eta \hspace{5 pt} \rm{s^{-1}}
\end{eqnarray}
The stiffness $\omega_0^2$ is mainly due to the gravitational pull
by the outer core and, as seen in equation (\ref{eq:omega_0^2}),
determined by the density difference in the core with the
gravitational constant, $G = 6.67 \times 10^{-11}$ N m$^2$
kg$^{-2}$. The stiffness due to Sun's tidal force that acts to
enlarge any displacements of the inner core from the center of the
Earth twice per its revolution is about 8 orders of magnitude
smaller than the gravitational stiffness and ignored. The stiffness
due to the Moon's tidal force is about twice larger than the one due
to the Sun's tidal force and ignored. The stiffness due to the
centrifugal force caused by Earth's revolution ($\approx$
$\omega_{R}^2$) is about two orders of magnitude smaller than the
gravitational stiffness and ignored.

$\Delta a$ in equation (\ref{eq:eq_of_motion}) is the magnitude of
the putative differential acceleration due to a violation of the
universality of free-fall in the Sun's gravitational field. The
signals of the violation are expected to have the frequency of once
per revolution of the Earth: $\omega_R$ $\approx$ 2$\pi$(24\
h)$^{-1}$ $\approx$ 7.3 $\times$ 10$^{-5}$ s$^{-1}$. At this
frequency, the major obstacle would be the tides (the body tides and
the ocean tides) which have the same frequency as the violation
signals. The tidal effects and other known effects can be largely
removed from the gravity data by applying appropriate models and
data analysis methods (e.g. \cite{Neumeyer2005}).

\section{\label{st:sensitivity}Expected sensitivity}

Assuming that $k < \omega_0$ (or $\eta < 5.9 \times 10^{12}$ Pa s),
the solution of equation (\ref{eq:eq_of_motion}) can be written by a
sum of two terms that express damped oscillations and forced
oscillations. The damped oscillations decay exponentially with time
$t$ ($e^{-kt}$; $k^{-1}$ is about 24 minutes for $\eta$ = 5.9
$\times$ 10$^{12}$ Pa s or when $k = \omega_0$), while the forced
oscillations remain:
\begin{equation}
x = A \sin(\omega_R t - \delta)
\end{equation}
where
\begin{eqnarray}
A &=& \frac{\Delta a}{\sqrt{(\omega_0^2 - \omega_R^2)^2 + 4 k^2
\omega_R^2}}\\
\tan \delta &=& \frac{2k \omega_R}{\omega_0^2 - \omega_R^2}
\end{eqnarray}
When the damping coefficient is sufficiently small (i.e. 4$k^2$
$\ll$ $\frac{\omega_0^4}{\omega_R^2}$ $\sim$ 4 $\times$ 10$^{-5}$
s$^{-2}$ or $\eta$ $\ll$ 2.8 $\times$ 10$^{13}$ Pa s), we obtain
\begin{equation}
A_{max} \approx \frac{\Delta a}{\omega_0^2}\label{eq:A}
\end{equation}
and $\tan\delta$ $\approx$ 0. When the inner core is shifted from
the center of the Earth by $A_{max}$, along the $x$-axis, this
displacement will change the gravitational field at a surface point
on the $x$-axis by
\begin{equation}
\Delta g \approx \frac{2Gm_{ic}}{r_e^3}\frac{\Delta a}{\omega_0^2}
\sim \frac{\Delta a}{10}\label{eq:Delta_g}
\end{equation}
where $r_e$ = 6.4 $\times$ 10$^{6}$ m is the average radius of the
Earth.

The noise level of superconducting gravimeters at quiet sites is a
few 10$^{-11}$ m s$^{-2}$ at the signal frequency \cite{Rosat2004}.
Assuming a somewhat better sensitivity of 10$^{-12}$ m s$^{-2}$ at
the signal frequency, we could estimate the universality of
free-fall to
\begin{equation}
\frac{\Delta a}{a_{sun}} \sim 1.6 \times
10^{-9}\label{eq:expected-sensitivity}
\end{equation}
where $a_{sun}$ $\approx$ 5.9 $\times$ 10$^{-3}$ m s$^{-2}$ is the
Sun's gravitational acceleration that acts on the Earth. Because of
the inclination of the Earth's rotation axis, the maximum violation
signals can be expected at observatories located on the equator in
Spring and Autumnal equinox points, and on Tropic of Cancer or
Capricorn in Summer and Winter solstices.

Equations (\ref{eq:A}) and (\ref{eq:Delta_g}) indicate that the
relative displacement of the inner core can be measured to $\sim$ 20
$\mu$m. For this amplitude, the Reynolds number is unity when $\eta$
$\simeq$ 200 Pa s. As a result, the range of the effective
viscosity, discussed above, is 200 Pa s $\leq$ $\eta$ $<$ 6 $\times$
10$^{12}$ Pa s. This range is consistent with the estimates
summarized in \cite{Secco1995}.

\section{\label{st:Discussion}Discussion}
This work is based on the simple Earth model and configuration.
Elaborate modellings have to be applied for more accurate analyses
of expected violation signals and of the expected sensitivity.
Especially, the value of $\omega_0$, which affects the estimates of
the expected signals and sensitivity, is highly model dependent. The
latest theoretical studies with elaborate Earth models predict
smaller values of $\omega_0$ than the value we have used (equation
(\ref{eq:omega_0^2})), between $\sim$ 2$\pi$(6 h)$^{-1}$ and $\sim$
2$\pi$(4 h)$^{-1}$ \cite{Rieutord2002, Rogister2003}. Because the
expected sensitivity is proportional to $\sim \omega_0^{-2}$, the
expected sensitivity could be better by $\sim$ 3 to 6 times by
applying elaborate Earth models. However, we have assumed the rather
optimistic sensitivity of 10$^{-12}$ m s$^{-2}$. Therefore, the
expected sensitivity would remain to be on the order of 10$^{-9}$.

The current limits on violations of the universality of free-fall
are on the order of 10$^{-13}$. In order to get comparable results,
a sensitivity better than a few 10$^{-15}$ m s$^{-2}$ (0.1 picogal)
at the signal frequency is required. Presently, superconducting
gravimeters are the most sensitive instruments for measurements in
the frequency range. Though it seems difficult to achieve this
sensitivity, there are several possibilities to improve the
sensitivity. One way may be applications of more sophisticated data
analyses than the usual Fourier analyses, as discussed in
\cite{Rosat2005} to search for the Slichter triplet. Another way may
be to carry out coincidence measurements with two superconducting
gravimeters located ideally opposite sides of the Earth near the
equator. If there were a violation towards the Sun, the expected
magnitude of the violation signal at the two superconducting
gravimeters is the same but the sign should be opposite. By
combining such coincidence signals, we could double the magnitude of
the expected signals and the sensitivity would be improved by a
factor of two. Candidate observatories for such coincidence
measurements may be the one at Hsinchu in Taiwan (25$^{\circ}$N
121$^{\circ}$E), where we are currently setting up two new
superconducting gravimeters, and the one at Concepcion in Chile
(37$^{\circ}$S 73$^{\circ}$ W). If the noise level of data from the
observatories were high, it might be better to use data from low
noise sites considering the degrees of signal compensation depending
on the latitude and longitude of the sites. Such global observations
would be possible through the Global Geodynamics Project network
(GGP \cite{GGP}). Further studies are necessary to figure out the
optimal schemes for global observations and noise reduction.
Currently, we are checking the performance of the new
superconducting gravimeters, installed in Hsinchu Taiwan in March
2006. We plan to carry out a preliminary test of the universality in
the near future.

As described in section \ref{st:Concept}, the difference in the mass
fraction due to gravitational self-energy is $\alpha_{grav}$ $\sim$
4 $\times$ 10$^{-10}$. Therefore, from equation
(\ref{eq:expected-sensitivity}) and the above discussion, the
geophysical test with current superconducting gravimeters would be
only sensitive to the chemical composition-dependent effect, but not
sensitive enough to test the universality of gravitational
self-energy. If the sensitivity were improved to $\sim$ 0.1 picogal,
we could test the universality of free-fall of gravitational
self-energy to the same level as the LLR experiment
\cite{Baessler1999,Adelberger2001,Williams2004}.

\section{Conclusions}

We have considered a new method of testing the universality of
free-fall. In this method, differential acceleration between the
Earth's inner core and the rest part of the Earth is to be searched
for by measuring surface-gravity effects. Based on a simple model,
we have shown that the universality would be tested to a level of
10$^{-9}$ with current sensitive superconducting gravimeters. Some
improvements can be expected from combinations of global
measurements and developments of methods of data analysis. We plan
to carry out a preliminary test of the universality using
superconducting gravimeters in Hsinchu Taiwan. To get a comparable
result with the LLR experiment, the sensitivity has to be improved
by about four orders of magnitude at the signal frequency (once per
day). A breakthrough in developments of gravity measurements is
necessary to achieve this sensitivity.

\begin{acknowledgments}
The author would like to thank H. Hatanaka, J.-S. Wu, S.-S. Pan,
C.-W. Lee, C. Hwang, B. F. Chao, and W. Kuang for their useful
discussions and comments on the manuscript. The author also wishes
to acknowledge C. C. Speake, C. Trenkel, C.-C. Chang and A. Pulido
Pat\'{o}n for their helpful discussions on the motion of the inner
core in the early phases of this work. The observatory in Hsinchu is
sponsored by the Ministry of Interior of Taiwan, and operated by
NCTU and ITRI.
\end{acknowledgments}

\bibliography{apssamp}

\end{document}